\documentclass{article}
\usepackage{amssymb}
\usepackage{amsmath,bm}
\usepackage{array}
\usepackage{caption,graphicx}
\usepackage{xcolor}

\newcommand{\be}{\begin{equation}}
\newcommand{\ee}{\end{equation}}
\def\calh{{\mathcal H}}
\def\Aut{{\mbox{Aut}}}
\def\calc{{\mathcal C}}
\def\cals{{\mathcal S}}
\def\un{{\rm 1\mkern-4mu I}}

\title{\bf Superconnection in the spin factor approach
 to particle physics}

\author{Michel DUBOIS-VIOLETTE$^{1)}$, Ivan TODOROV$^{2)}$}
\date{
	\small 
	$^{1)}$ Laboratoire de Physique des 2 Infinis Ir\`ene Joliot Curie\\
	P\^ole Th\'eorie, 
	IJCLab UMR 9012\\
	CNRS, Universit\'e Paris-Saclay, 
	B\^atiment 210\\
	F-91406 Orsay Cedex, France\\
	michel.dubois-violette@u-psud.fr
	\bigskip
	\\ $^{2)}$ Institut des Hautes \'Etudes Scientifiques, 35 route de Chartres, 
	\\ F-91440 Bures-sur-Yvette, France
	\\ and
	\\ Institute for Nuclear Research and Nuclear Energy, Bulgarian Academy of Sciences 
	\\ Tsarigradsko Chaussee 72, BG-1784 Sofia, Bulgaria
	\\ (permanent address)
	\\ ivbortodorov@gmail.com}
	
\begin{document}
	\maketitle
	\thispagestyle{empty}
	\vglue 1,5cm

\begin{abstract}
The notion of superconnection devised by Quillen in 1985 and used in gauge-Higgs field theory in the 1990's is applied to the spin factors (finite-dimensional euclidean Jordan algebras)  recently considered as representing the finite quantum geometry of one generation of fermions in the Standard Model of particle physics.
\end{abstract}
\bigskip

\newpage

\tableofcontents

\bigskip

\noindent {\bf References}

\newpage

\section{Introduction}\label{sec1}

It is natural to expect that the finite spectrum of fundamental particles of matter corresponds to representations of a finite-dimensional algebra of quantum observables endowed with some further structure. On the basis of the spectral theory needed for quantum mechanics, these finite-dimensional algebras of quantum observables have been identified as the finite-dimensional euclidean (or formally real) Jordan algebras \cite{jor:1932}, \cite{jor:1933} and have been classified 
\cite{jor-vn-wig:1934}. These algebras are the quantum analogues of the finite-dimensional algebras of real functions,  it is convenient to consider them as algebras of ``real functions" on virtual ``finite quantum spaces". We will use freely this analogy by refering to ``the finite quantum space" corresponding to a finite-dimensional euclidean Jordan algebra. Any finite-dimensional euclidean Jordan algebra has a unit and is the direct sum of a finite number of simple ideals and the simple finite-dimensional euclidean Jordan algebras fall into 3 classes~ :
\begin{enumerate}
\item
The hermitian $n\times n$-matrices $J^1_n=\calh_n(\mathbb R)$, $J^2_n=\calh_n(\mathbb C)$ and $J^4_n=\calh_n(\mathbb H)$ over the reals, the complexes and the quaternions, for $n\geq 3$ and \linebreak[4] $\mathbb R\ (=\calh_1(\mathbb R)=\calh_1(\mathbb C)=\calh_1(\mathbb H)) $.

\item
The spin factors $J^n_2=JSpin_{n+1}$ $(n\geq 1)$.

\item
The exceptional Jordan algebra of hermitian $3\times 3$-matrices $J^8_3=\calh_3(\mathbb O)$ over the octonions.
\end{enumerate}

The Jordan algebra $J^8_3$ is exceptional in the sense that it cannot be realized as a subspace of an associative algebra stable under the symmetrized product  \cite{alb:1934}. The classes 1 and 2 contain only special (i.e. non exceptional) Jordan algebras. However there is an important difference between Class 1 and Class 2. Namely the Jordan algebras which belong to Class 1 are the real subspaces of all hermitian elements of associative $\ast$-algebras while in the case of Class 2  the spin factors $JSpin_n$ are only Jordan subalgebras of the Jordan algebras of all hermitian elements of associative $\ast$-algebras. This fact is in particular relevant for the spin factor $J_2^8=JSpin_9=\calh_2(\mathbb O)$ which in our approach corresponds to the finite quantum geometry of one generation of the standard model. This is the very reason of the existence of the euclidean extension $\tilde J_2^8$ of $J^8_2$ which contains the internal observables of the fundamental particles of matter for one generation.\\

The Jordan algebra approach to the finite quantum geometry of particle physics models was originally developed \cite{mdv:2016b},  \cite{mdv-tod:2019} in the context of the exceptional Jordan algebra $J_3^8 = {\mathcal H}_3 ({\mathbb O})$. It was realized in  \cite{tod-mdv:2017}, \cite{tod-dre:2018},  \cite{mdv-tod:2019} that the quantum geometry of one generation is captured by a special Jordan algebra -- the 10-dimensional spin factor
\be
\label{eq11}
J_2^8 = JSpin_9 = {\mathcal H}_2 ({\mathbb O}) (\subset J_3^8) \, ,
\ee
i.e. the $2 \times 2$ hermitian matrices with octonionic entries. The gauge symmetry group of the Standard Model (SM) of particle physics,
\be
\label{eq12}
G_{\rm SM} = S(U(3) \times U(2)) = \frac{SU(3) \times SU(2) \times U(1)}{{\mathbb Z}_6}
\ee
is the subgroup of the automorphism group ${\rm Spin} (9)$ of $J_2^8$ that preserves the splitting
\be
\label{eq13}
{\mathbb O} = {\mathbb C} \oplus {\mathbb C}^3
\ee
and acts $\mathbb C$-linearly on $\mathbb C^3$.\\

 The splitting (\ref{eq13}) is preserved by the subgroup $SU(3)$ of the automorphism group $G_2$ of the octonions which was identified long ago to the colour symmetry of quarks by G\"ursey and G\"unaydin  \cite{gun-gur:1974},  \cite{gur:1974}. From the point of view of physics (\ref{eq13}) corresponds to the quark-lepton symmetry.  Conversely, it was shown in \cite{mdv:2016b} that  the unitarity and the unimodularity of $SU(3)$ lead directly to a unital algebra structure on $\mathbb C\oplus \mathbb C^3$ which is isomorphic to $\mathbb O$ as real algebra, $SU(3)$ being then the group of $\mathbb C$-linear automorphisms. In other words, this associates the quark-lepton symmetry to the unimodularity of the colour group and selects the euclidean Jordan algebras $J^8_2=\calh_2(\mathbb O)$ and $J^8_3=\calh_3(\mathbb O)$ endowed with their automorphisms preserving the splitting (1.3), (notice that $\calh_1(\mathbb O)=\mathbb R$ and that the $\calh_n(\mathbb O)$ for $n\geq 4$ are not Jordan algebras).\\

 The resulting characterization of $G_{\rm SM}$ was recently commented in   \cite{kra:2019} where the action of ${\rm Spin} (9)$ on a pair of octonions (that spans the spinor representation ${\bf 16} \simeq {\mathbb O}^2$ of ${\rm Spin} (9)$ and appear in the 27-dimensional algebra $J_3^8$) is exploited.  A Jordan algebra modification of Connes' non-commutative geometry approach to the SM,\cite{ac-lot:1990}, \cite{cha-ac:2010} is developed in \cite{boy-far:2019}.

\smallskip

There are three Lie algebras associated with the Jordan algebra $J^8_2$ (1.1):
\be 
\label{3Lie}
Der(J^8_2) = so(9) \subset str(J^8_2) =so(9, 1) \oplus \mathbb R_+ \subset co(J^8_2) = so(10, 2).
\ee
Here $so(9)$ is the \textit{Lie algebra of derivations} (infinitesimal automorphisms) of $J^8_2$; the \textit{structure Lie algebra} $str(J^8_2)$ is the derivation algebra of the positive cone $\mathcal C = \mathcal C(J^8_2)$ of states (invertible elements of $J^8_2$ that may be written as sums of squares); $so(10,2)$ is the \textit{conformal Lie algebra} of $J^8_2$ that leaves invariant the tube domain $J^8_2 + i\mathcal C$. We shall also use in what follows the Lie algebra $so(10)$ of the popular Grand Unified Theory (GUT), which appears here as the compact real form of the complexification of $so(9, 1)$ but also as the semi-simple part of the maximal compact Lie subalgebra $so(10)\oplus so(2)$ of $co(J^8_2)$. 

In the present paper we observe that the Quillen's notion of superconnection \cite{qui:1985}, \cite{mat-qui:1986} readily applies to theories based on the Clifford algebra $C\ell(4n+1, 1), n=1, 2, ...$, and we work out the bosonic mass relations (and the associated Weinberg angle) for $n=2$ applying the superconnection approach to the euclidean extension of $J^8_2$ with an $so(10)$ symmetry. (In earlier applications of  superconnections to particle physics - see \cite{coq:1991}, \cite{nee-ste:1990},  \cite{roe:1999}, \cite{ayd-min-sun-tak:2015} and references therein - the bosonic mass relations and the Weinberg angle have only been computed for the  $U(2)$ electroweak model, albeit in \cite{nee-ste:1990} all quark and lepton quantum numbers have been fitted in a representation of the $sl(5|1)$ Lie superalgebra.)

\smallskip

As noted in \cite{mdv:2016b} and elaborated in  \cite{kra:2019} one can similarly derive the electroweak subgroup $U(2)$ of the gauge group $G_{\rm SM}$ of the SM from the automorphism group ${\rm Spin} (5)$ of the spin factor $J_2^4 = {\mathcal H}_2 ({\mathbb H})$:
\be
\label{eq15}
{\rm Aut} (J_2^4) = {\rm Spin} (5) = U(2,{\mathbb H}) \, , \quad J_2^4 = {\mathcal H}_2 ({\mathbb H})
\ee
with the alternative (but non-associative) ring ${\mathbb O}$ of octonions substituted by the associative division algebra ${\mathbb H}$ of quaternions. The superconnection approach applies equally well to the ``mini internal space'' $J_2^4$ of the electroweak model of leptons which can thus serve as a simpler ``toy model'' for $J_2^8$. 

\smallskip

We begin in Sect.~2 by summarizing our treatment of the euclidean extension of $J_2^8$, introduced in [16], 
\be
\label{eq16}
\widetilde J_2^8 := {\mathcal H}_{16} ({\mathbb C}) \oplus {\mathcal H}_{16} ({\mathbb C}) = J_{16}^2 \oplus J_{16}^2
\ee
(that admits an analogue $\widetilde J_2^4$ for $J_2^4$). We recall the notion of $U(n)$ superconnection expounded by Roepstorff [30] and define its extension to $c\ell(4n+1, 1)$. In Sect.~3 we recall the fermionic oscillator realization of $C\ell(9, 1)$ and characterize the 16-dimensional \textit{particle subalgebra} $J(\mathcal{P})$ of $\widetilde J_2^8$. In Sect. 4 we introduce the Higgs potential allowing a symmetry breaking minimum and derive the mass matrix for the gauge fields. Section 5 is our temporary conclusion.\\

Our notations and conventions are the ones of \cite{mdv:2016b} and \cite{mdv-tod:2019} and of 
\cite{tod:2019} in particular for the Clifford algebras and their ``fermionic oscillator" (or Canonical Anticommutation Relations) representations for the even-dimensional case. Concerning the latter point, it should be mentioned that the representation of the Clifford algebra of an even-dimensional euclidean space as the CAR algebra depends on the choice of a direction of  simple spinor in the sense of Elie Cartan which is the corresponding direction of the Fock vacuum \cite{mdv:1993}. In fact the directions of simple spinors parametrize the isometric complex structures (see also \cite{mdv:2013} for a more general point of view). It is worth noting  that Sections 3.2 and 3.3 of \cite{mdv:2016b} and Section 2 of \cite{tod:2019} contain motivated summaries of the Jordan-von Neumann-Wigner classification and that, in this respect, \cite{tod:2019} is a fairly complete reference. For Jordan algebras and Jordan modules our reference is \cite{jac:1968} and for exceptional Lie groups see 
\cite{yok:2009}.

%%%
\section{Internal symmetry and superconnection} \label{sec2}
\setcounter{equation}{0}

As explained in Sect.~4 of \cite{mdv-tod:2019} and in Sect.~2.2 of \cite{tod:2019} the optimal euclidean extension of $J_2^8$ is the direct sum (\ref{eq16}) of two Jordan algebras of complex hermitian $16 \times 16$ matrices. It contains, in particular, the hermitean generators $i \, \Gamma_{ab}$, $a,b = 0,1,\ldots , 8$ of the derivation algebra $so(9)$ viewed as a sub Lie algebra of $so(9,1) \subset C\ell^0 (9,1) \simeq {C\ell}(9, 0)$, the (restricted) structure algebra of $J_2^8$. Choosing a basis $(e_0 = 1 , e_1 , \ldots , e_7)$ of octonion units we can think of $J_2^8$ as generated by the $2 \times 2$ hermitian octonionic matrices
$$
\widehat e_a = \begin{pmatrix} 0 &e_a \\ e_a^* &0 \end{pmatrix} , \quad a = 0,1,\ldots , 7 \quad (e_0^* = e_0 , e_j^* = -e_j \ \mbox{for} \ j = 1, \ldots , 7) , 
$$
\be
\label{eq21}
\widehat e_8 = \sigma_3 = \begin{pmatrix} 1 &0 \\ 0 &-1 \end{pmatrix} .
\ee
We shall represent $\hat{e}_a$ by the products
$$
\Gamma_{-1} \Gamma_a , \quad a = 0,1,\ldots , 8, \quad [\Gamma_a , \Gamma_b]_+ := \Gamma_a \Gamma_b + \Gamma_b \Gamma_a = 2 \delta_{ab} \, ,
$$
\be
\label{eq22}
[\Gamma_{-1} , \Gamma_a]_+ = 0 \, , \quad \Gamma_{-1}^2 = - \un
 \Rightarrow (\Gamma_{-1} \Gamma_a)^2 = \un \, ,
\ee
where $\Gamma_{\alpha}$, $\alpha = -1,0,\ldots , 8$, generate the Clifford algebra ${C\ell} (9,1)$. The Coxeter element $\omega_{9,1}$ of ${C\ell} (9,1)$ plays the role of chirality and commutes with $so(9,1)$:
\be
\label{eq23}
\gamma := \omega_{9,1} = \Gamma_{-1} \Gamma_0 \Gamma_1 \ldots \Gamma_7 \Gamma_8 \, , \ \gamma^2 = \un
 \, ; \ [\gamma , \Gamma_{\alpha\beta}] = 0 \ \mbox{for} \ \Gamma_{\alpha\beta} = \frac12 [\Gamma_{\alpha} , \Gamma_{\beta}] \, .
\ee
In a representation in which $\gamma = \sigma_3\otimes \textbf{1}_{16}$ the 32-dimensional Dirac spinor representation of $so(9,1)$, generated by $\Gamma_{\alpha\beta}$, is reduced:
\be
\label{eq24}
{\bf 32} = {\bf 16}_L \oplus {\bf 16}_R \, , \ (\gamma - 1) {\bf 16}_L = 0 = (\gamma + 1) {\bf 16}_R \, .
\ee
The ${C\ell}(9,1)$ generators anticommute with chirality and intertwine left and right chiral (Weyl) spinors
\be
\label{eq25}
[\Gamma_{\alpha} , \gamma]_+ = 0 \, , \ \Gamma_{\alpha} : {\bf 16}_{L,R} \to {\bf 16}_{R,L} \, , \ \alpha = -1,0,1,\ldots , 8 \, .
\ee
In Haag's approach \cite{haa:1993} to quantum field theory the algebra of observables is a subalgebra of gauge invariant elements (with respect to the unbroken gauge symmetry) of a larger {\it field algebra}. We shall postulate that the finite-dimensional (internal space) counterpart of the field algebra is  ${\mathbb Z}_2$-graded complex Clifford algebra 
$C\ell_{10} = C\ell (10, \mathbb C)$ whose algebra of derivations 
$so(10, \mathbb C)$ belongs to its even part. The matrices 
$\Gamma_{\alpha \beta}$ (2.3)provide an orthonormal (with respect to the trace product) basis of the Lie algebra $so(9, 1)$. The corresponding hermitian matrices 
\be
\label{eq26}
\Gamma_{-1a} , i\Gamma_{ab} \in so(10,{\mathbb C}) (\supset so(9,1)) \, , \ a,b = 0,1,\ldots , 7,8, 
\ee
belong to the exteded observable algebra $\widetilde J_2^8$ (1.6) and form a basis of $i so(10) \, (i=\sqrt{-1})$ . The odd part of $C\ell_{10}$ (that anticommutes with $\gamma)$) includes its hermitian generators $(i\Gamma_{-1}, \Gamma_a, a=0, ..., 8)$ and will give room to the (Lorentzian scalar) {\it Higgs fields}.

\smallskip
We proceed to identifying the symmetry generators and a complete set of commuting observables.
\smallskip
Singling out $e_7 \in {\mathbb O}$ as the imaginary unit preserved by $SU(3)$ we can write the decomposition (\ref{eq13}) in the form (cf. Appendix):
$$
{\mathbb O} \ni x = z + Z \, , \ z = x^0 + x^7 e_7 \, , \ Z = Z^1 e_1 + Z^2 e_2 + Z^4 e_4 \, ,
$$
\be
\label{eq27}
Z^j = x^j + x^{3j({\rm mod} \, 7)} e_7 \, , \ j = 1,2,4,
\ee
where we have used the octonionic multiplication rules of \cite{bae:2002}
\be
\label{eq28}
e_i \, e_{i+1} = e_{i+3({\rm mod} \, 7)} (= -e_{i+1} \, e_i) \, , \ i = 1,2,\ldots , 7 \, .
\ee
It is intriguing to observe that the Lie subalgebra of $so(9, 1)$ which preserves the splitting (2.7) and acts complex linearly on $Z$ is $so(3,1)\oplus u(3)$. One could be tempted to connect it with two unbroken symmetries, the Lorentzian $so(3, 1)$ and the colour $su(3)=su(3)_c$, plus a $u(1)$ whose generator will be identified with $B-L$. (Its conservation may be broken by a Majorana mass term.) We, however, will continue to follow here our philosophy and will interpret the finite quantum algebra in terms of internal degrees of freedom; consequently we shall only use the compact real form $so(10)$ of the complexified $so(9, 1)$.
  
 The Pati-Salam Lie subalgebra
\be 
\label{PatiSalam}
su(2)_L \oplus su(2)_R \oplus su(4) \subset so(10)
\ee
appears as the maximal Lie subalgebra of $so(10)$ that preserves the splitting of the vector representation $\textbf{10}=\textbf{6}+\textbf{4}$ (but not the quark lepton splitting in the spinor representation $\textbf{16}$). It is realized as follows in terms of the matrices $\Gamma_{\alpha \beta}$: 
\begin{eqnarray} 
\label{P-Sgenerators}
su(4) \simeq so(6) = {\rm Span} \{ \Gamma_{jk} , \ j,k = 1,2,\ldots , 6 \} \, , \nonumber \\
su(2) \oplus su(2) \simeq so(4) = {\rm Span} \{i\Gamma_{-1 \alpha}, \Gamma_{\alpha \beta}, \, \alpha,  \beta = 0,7,8\} \, .
\end{eqnarray}
In particular, we choose a basis of $su(3) \oplus u(1)$ invariant commuting observables:
\be
\label{eq210}
2I_3^L = \frac12 (\Gamma_{8-1} - i \Gamma_{07}) \, , \ 2I_3^R = -\frac12 (\Gamma_{8-1} + i \Gamma_{07}) \Rightarrow I_3^L I_3^R = 0 \, ,
\ee
\be
\label{eq211}
B-L = \frac i3 (\Gamma_{13} + \Gamma_{26} + \Gamma_{45}) \, ,
\ee
$B$ and $L$ being the baryon and the lepton numbers. The colour gauge Lie algebra $su(3)_c$ then appears as the commutant of $B-L$ in $su(4)$. The weak hypercharge $Y$ and the electric charge $Q$ are expressed as:
\be
\label{eq212}
Y = B - L + 2 I_3^R \, , \ Q = I_3^L + \frac12 Y = \frac12 (B-L) + I_3, \, I_3:= I_3^L+I_3^R=-\frac i2 \Gamma_{07} \, .
\ee
The left and right isospins take values $0$ and $1/2$ so that $2I_3^L$ and $2I_3^R$ satisfy
\be
\label{eq213}
(2I_3^X)^3 = 2I_3^X \quad \mbox{for} \quad X=L,R \Rightarrow  P_1 : = (2I_3^L)^2 = P_1^2 = \un 
- (2I_3^R)^2 \, .
\ee
($P_1$ being the $SU(2)_L$ invariant projector on the states of weak isospin $1/2$.)

\smallskip

We can write the (skew hermitian) matrix valued gauge field 1-form as:
\be
\label{eq214}
\widehat A = dx^{\mu} A_{\mu}^s X_s = i \widehat W + i \widehat B + i \widehat G
\ee
where $s = 1, \ldots , 12 = \dim G_{\rm SM}$ and $X_s$ are suitable linear combination of the matrices (\ref{eq26}); the three terms $\widehat W , \widehat B , \widehat G$ correspond to the subalgebras $su(2)_L$, $u(1)_Y$, $su(3)_c$, respectively, of the Lie algebra
\be
\label{eq215}
{\mathcal G}_{\rm SM} = su(2)_L \oplus u(1)_Y \oplus su(3)_c
\ee
of the gauge group $G_{\rm SM}$ (\ref{eq12}); they will be displayed explicitly in Sect. 3 below.

We shall interrupt for a moment our exposition in order to summarize, for reader's convenience, the notion of a superconnection on the example of the gauge group $U(n)$ acting on the exterior algebra $\bigwedge \mathbb{C}^n$ as worked out in \cite{roe:1999}. We shall identify the $\mathbb{Z}_2$ grading of $\bigwedge \mathbb{C}^n$ with chirality, assuming (arbitrarily) that $\bigwedge^0 \mathbb{C}^n$ is right chiral (i.e. has negative chitality) and denote by $A^\pm$ left and right chiral propjections of the $U(n)$ connection $\hat{A}$. We then define the $U(n)$ \textit{superconnection} 1-form on $T^*M\otimes\bigwedge \mathbb{C}^n$ by    
\be
\label{eq217}
{\mathbb D} = d + \widehat A + \widehat\Phi \, , \ \widehat A = \begin{pmatrix} A^+ &0 \\ 0 &A^- \end{pmatrix} , \ \widehat\Phi = \begin{pmatrix} 0 &\phi^* \\ \phi &0 \end{pmatrix}
\ee
where $d = dx^{\mu} \partial_{\mu}$ and the two by two block matrix has $2^{n-1} \times 2^{n-1}$ dimensional blocks. The ${\mathbb Z}_2$ grading of $1$-forms is the combined grading of fields (in which $A_{\mu}^+$ and $A_{\mu}^-$ are even and $\widehat\Phi$, $\phi^*$, $\phi$ are odd) and of differential forms (in which $dx^{\mu}$ is odd, $dx^{\mu} \wedge dx^{\nu}$ is even, etc.). Thus the superconnection ${\mathbb D}$ is odd. The corresponding curvature form is obtained using the ${\mathbb Z}_2$ graded commutator:
\be
\label{eq218}
{\mathbb F} = \widehat F + {\mathbb D} \, \widehat\Phi \, , \quad \widehat F = DA \, , \ {\mathbb D} \, \widehat\Phi = [{\mathbb D} , \widehat\Phi]_+
\ee
where $D\widehat A = dx^{\mu} \wedge dx^{\nu} \begin{pmatrix} F_{\mu\nu}^+ &0 \\ 0 &F_{\mu\nu}^- \end{pmatrix}$, $F_{\mu\nu}^\pm = \partial_{\mu} A_{\nu}^{\pm} - \partial_{\nu} A_{\mu}^{\pm}$, while
$$
[{\mathbb D} , \widehat\Phi]_+ = \widehat\Phi^2 + \begin{pmatrix} 0 &(D\phi)^* \\ D\phi &0 \end{pmatrix} , \ D\phi = D^- \phi + \phi D^+ = dx^{\mu} ((\partial_{\mu} + A_{\mu}^-) \phi - \phi A_{\mu}^+) ,
$$
\be
\label{eq219}
(D\phi)^* = dx^{\mu} ((\partial_{\mu} + A_{\mu}^+) \phi^* - \phi^* A_{\mu}^-) \, .
\ee
In the last two equations we have used the anticommutativity of $\phi^{(*)}$ and $dx^{\mu}$.
We observe that the above construction works once one has the notion of chirality which allows to define "the Higgs" as a matrix valued chirality changing scalar field. Remarkably, embedding our $\tilde{J}_2^8$ model into $C\ell(9, 1)$ provides a natural notion of chirality, Eq. (\ref{eq23}), such that the operator 
\be 
\label{eq220}
\widehat\Phi = \phi^{\alpha} \Gamma_{\alpha}  
\ee 
is chirality changing. For $\gamma = \sigma_3\otimes \textbf{1}_{16}$ the matrices (\ref{eq219}) are reproduced. 
\smallskip
\section{Fermionic oscillators. Particle subspace}\label{sec3} 
\setcounter{equation}{0}
We shall use the following fermionic oscillators' representation of ${C\ell} (10,{\mathbb C})$ (cf. \cite{fur:2018}, \cite{tod-dre:2018}, \cite{tod:2019}):
$$
2a_0 = \Gamma_0 + i\Gamma_7 \, , \ 2a_j = \Gamma_1 + i\Gamma_3j(mod 7), j=1,2,4, 2a_8 = \Gamma_8 + \Gamma_{-1}$$
$$
(2a_0^* = \Gamma_0 -i\Gamma_7, 2a_1^* = \Gamma_1 -i\Gamma_3, ..., 2a_8^* = \Gamma_8 - \Gamma_{-1}),$$
\be
\label{a(*)}
[a_{\mu} , a_{\nu}]_+ = 0 \, ,  [a_{\mu} , a_{\nu}^*]_+ = 2 \delta_{\mu\nu} \, \mu, \nu = 0, 1, 2, 4, 8.
\ee
The basic fermions and antifermions are given by the primitive idempotents of the abelian (unital) algebra generated by the Cartan subalgebra of the (complexified) $so(9, 1)$. It is spanned by the idempotents
\be 
\label{pi's} 
\pi_{\nu} = a_{\nu} a_{\nu}^*(=\pi_{\nu}^2), \pi_{\nu}^{\prime}=a_{\nu}^*a_{\nu}=1-\pi_{\nu} \, (\pi_\nu\pi_\nu^{\prime}=0), \nu=0,1,2,4,8.
\ee
They belong to the euclidean extension $\tilde{J}_2^8$ (\ref{eq16}) of the octonionic spin factor $J_2^8$. We postulate that the symmetry algebra is invariant under separate phase transformations of the quark and lepton oscillators: $a_\mu \rightarrow e^{i\alpha}a_\mu (a_\mu^*\rightarrow e^{-i\alpha}a_\mu^*, \, \mu =0, 8),  a_j\rightarrow e^{-i\beta} a_j, (a_j^*\rightarrow e^{i\beta} a_j^*, \, j=1, 2, 4)$, or equivalently, it commutes with:
\be 
\label{IR_B-L}
2I_3^R=\frac{1}{2}([a_0, a_0^*]+[a_8, a_8^*]), \, B-L=\frac{1}{3}\sum_{j=1,2,4}[a_j^*, a_j].
\ee
The resulting \textit{symmetry subalgebra} of the Pati-Salam Lie algebra (\ref{PatiSalam}) (which also preserves the complex linearity of the $su(3)_c$ action) is the $u(1)$ extension
\begin{eqnarray}
\label{g-ext}
\mathfrak{g}_{ext} = u(2) \oplus u(3), \, \, u(2) = Span\{a_{\mu}^* a_\nu, \mu, \nu = 0,8\}, \nonumber \\
u(3)= Span\{a_j^*a_k, j, k = 1, 2, 4\},  \,
\end{eqnarray}
of the gauge Lie algebra $\mathfrak{g}_{SM} = s(u(2)\oplus u(3))$ of the SM. In particular, the (left) electroweak $su(2)_L$ symmetry generators,
\be 
\label{su2L}
I_+^L= a_8^* a_0, \, I_-^L= a_0^*a_8, \, 2I_3^L= [I_+^L, I_-^L] = \pi_8^{\prime}- \pi_0^{\prime},
\ee
are complemented by $2I_3^R$ and $B-L$ (\ref{IR_B-L}) (cf (2.10) (2.11)). 
%2I_3^R= \pi_8-\pi_0^{\prime}, \, B-L = \frac{1}{3}\sum_{j=1,2,4}[a_j^*, a_j] = \frac{1}{3}\sum_j(\pi_j^{\prime} - \pi_j)
The $u(1)$ centre of $\mathfrak{g}_{SM}$ is spanned by the hypercharge
\be 
\label{hypercharge}
Y = B-L + 2I_3^R = \frac{2}{3}(\pi_1^{\prime} + \pi_2^{\prime} + \pi_4^{\prime}) - \pi_0^{\prime} - \pi_8^{\prime},
\ee 
the linear combination of $B-L$ and $2I_3^R$ that annihilates the right chiral (sterile) neutrino:
\be 
\label{sterileR}
(\nu_R): = |\nu_R><\nu_R| = \pi_0\pi_1\pi_2\pi_4\pi_8 \Rightarrow Y(\nu_R) = 0.
\ee 
A general problem in theories with configuration space of the form $\calc(M) \otimes {\mathcal F}$, the product of the commutative algebra of smooth functions on a spin manifold $M$ with a finite dimensional
(not necessarily commutative or associative) algebra ${\mathcal F}$, first encountered in the better developed noncommutative geometry approach \cite{ac-lot:1990}, \cite{cha-ac:2010},  is the problem of fermion doubling (or rather quadrupling) \cite{gra-ioc-sch:1998}, recently tackled in \cite{boc-sit:2020}. In order to avoid (or reduce) the problem one can simply restrict attention to the 16 dimensional {\it particle subalgebra}
\be
\label{eq2201}
J({\mathcal P}) = {\mathcal H}_8^L ({\mathbb C}) \oplus {\mathcal H}_8^R ({\mathbb C})
\ee
of the Jordan algebra (\ref{eq16}). The projector $\mathcal{P}$ on the particle subspace can be written as the sum of projectors $\ell$ and $q$ on the lepton and the quark subspaces:
\begin{eqnarray}
\label{Plq}
\mathcal{P}=\ell+q(=\mathcal{P}^2), \, \bar{\mathcal{P}} (= 1-\mathcal{P})= \bar{\ell} + \bar{q}, \, \ell = \pi_1\pi_2\pi_4 \, (L=\ell-\bar{\ell}),  \nonumber \\
\bar{\ell} = \pi_1' \pi_2' \pi_4', \, \, \, q = \sum_{j=1,2,4} U_j \bar{\ell}= \pi_1\pi_2'\pi_4' + \pi_1'\pi_2\pi_4'+\pi_1'\pi_2'\pi_4.
\end{eqnarray}
Here $U_j = U(a_j^*, a_j)$ is the (polarized) \textit{quadratic Jordan operator} (see  Eq. (3.24) of \cite{tod:2019}
and references cited there):
\be 
\label{Uaa8}
U_\nu X := a_\nu^*Xa_\nu + a_\nu Xa_\nu^*.
\ee
The gauge invariant states of the subalgebra $J(\mathcal{P})$  are uniquely characterized by the eigenvalues of $2I_3^L$ (\ref{eq210}) (3.5) and $Y$ (\ref{eq212}) (3.6). In particular, the chirality $\gamma$ in $J(P)$ is determined by  anyone of these quantum numbers:
\be
\label{eq221}
\gamma + (-1)^{2I_3^L} = 0 = \gamma + (-1)^{3Y} \, .
\ee
Conversely, Eq.~(\ref{eq221}) determines the subalgebra $J({\mathcal P})$. The orthogonal projector ${\mathcal P} : J_{16}^2 \oplus J_{16}^2 \to J({\mathcal P})$ is given by:
\be
\label{eq222}
{\mathcal P} (=\ell+q) = \frac12 (1-\gamma (-1)^{2I_3^L}) = \frac12 (1-\gamma (-1)^{3Y})  \, .
\ee
Clearly the projector $\mathcal{P}$ commutes with $\mathfrak{g}_{ext}$ so that $J(\mathcal{P})$ admits the same (extended) symmetry. The $SU(2)_L$-invariant projectors in $J({\mathcal P})$ are determined by the eigenvalues of $Y$. For the left chiral particles for which $P_1 = (2I_3^L)^2 = 1$ (cf. (\ref{eq213})) $Y$ takes two values, $-1$ and $\frac13$, of multiplicity two and six, respectively. In ${\mathcal H}_8^R ({\mathbb C})$, for $P_1=0$, the hypercharge takes four eigenvalues: two nondegenerate $Y = 0,-2$ and two others, $Y = \frac43 , -\frac23$ of multiplicity three each. We note that for the electroweak model (based on the Jordan algebra $J_2^4$ (\ref{eq15})) - with only leptons present - the trace of the hypercharge in the left and the right particle space is $-2$, so that only their difference, the \textit{supertrace}, vanishes (as emphasized in \cite{coq:1991}). By contrast, in the full SM the trace of $Y$ vanishes in $\mathcal{H}_8^L$ and $\mathcal{H}_8^R$, separately.

The expression (\ref{eq222}) for $\mathcal{P}$ together with the anticommutativity of $a_j^{(*)}(=a_j$ or $a_j^*)$ with $\gamma$ and their left isospin independence for $j=1, 2, 4$ implies that their projection on $J(\mathcal{P})$ vanishes:
\begin{equation}
\label{PajP}
a_j^{(*)}\gamma = -\gamma a_j^{(*)}, \, [I_3^L, a_j^{(*)}]=0 \Rightarrow \mathcal{P}a_j^{(*)}\mathcal{P}=0, \, j = 1, 2, 4.
\end{equation} 
 Thus, the projection of the Higgs field on the particle subspace commutes with the gluon field $G_\mu$ (which will be expressed in terms of $a_j^{(*)}$ in Eq. (\ref{AWBG}) below):
\begin{equation}
\label{HiggsInP}  
\hat{\Phi}(x) = \bar{\hat{\phi}}_0 a_0 + \phi_0 a_0^* + \bar{\phi}_8 a_8 + \phi_8 a_8^* \Rightarrow [\hat{\Phi}, G_\mu]=0.
\end{equation} 
Then the commutator $[A_\mu, \hat{\Phi}]$ appearing in the curvature form (cf. (\ref{eq218})), 
\be 
\label{DCliff}
\mathbb{D} = d+\hat{A}+\hat{\Phi}, \, \mathbb{D}^2 =
\hat{F} + d\hat{\Phi} +dx^\mu[A_\mu, \hat{\Phi}] + \hat{\Phi}^2, \, \hat{F} = d\hat{A} + \hat{A}^2,
\ee  
will only involve the electroweak fields which will acquire mass after the symmetry breaking. 
\smallskip

\section{Higgs potential and bosonic Lagrangian}\label{sec4}
\setcounter{equation}{0}
The bosonic action density in a gauge theory is defined as the trace of (half of) the product of the curvature with its Hodge star dual. In order to account for symmetry breaking we shall replace $\hat{\Phi}^4$ by a more general fourth order expression, invariant with respect to the unbroken gauge symmetry with Lie algebra
\be 
\label{unbroken} 
su(3)_c \oplus u(1)_Y \oplus u(1)_L \subset \mathfrak{g}_{SM}, \, \, u(1)_L = Span \{I_3^L = Q - \frac12 Y\}.
\ee 
(This extends the procedure adopted in \cite{roe:1999}
where one subtracts from $\hat{\Phi}^2$ a general $U(n)$ invariant operator.) We shall write the Higgs potential as:
\begin{eqnarray} 
\label{V(Phi)}
V(\phi) = \frac{1}{2}tr\left(\hat{\Phi}(\kappa P_1 + P_1^{\prime})\hat{\Phi}- m^2(P_1+\kappa P_1^{\prime})\right)^2 + \lambda \phi_0 \bar{\phi}_0\phi_8\bar{\phi_8}\nonumber \\ = \frac{1}{2}(\phi\bar{\phi}-m^2)^2 tr(P_1 + \kappa^2 P_1^{\prime}) + \lambda \phi_0 \bar{\phi}_0\phi_8\bar{\phi_8}, \, \, \, m, \kappa, \lambda >0.
\end{eqnarray}  
Here we have used the relations (cf. (\ref{eq213})):
\begin{eqnarray}
\label{PhiP1Phi}
P_1^{\prime}:=1-P_1 = (2I_3^R)^2 \, (P_1 P_1^{\prime} =0, P_1 + P_1^{\prime} = 1), \nonumber \\
\hat{\Phi}P_1 \hat{\Phi} =\phi\bar{\phi} P_1', \, \hat{\Phi}P_1'\hat{\Phi} =\phi\bar{\phi} P_1, \, \, 
\phi\bar{\phi} = \phi_0\bar{\phi}_0 + \phi_8\bar{\phi}_8  . 
\end{eqnarray}
It is the last, fourth order, term in (\ref{V(Phi)}) that breaks the $U(2)$ electroweak symmetry to $U(1)\times U(1)$ (the independent change of phases  of $\phi_0, \phi_8$). 
   
\smallskip
We shall write the bosonic Lagrangian of the SM in  the form:
\begin{eqnarray}
\label{LAphi}
L(A,\phi) &= &-\frac14 {\rm tr} (F_{\mu\nu} F^{\mu\nu}) + \frac12{\rm tr} (\partial_{\mu} \phi \partial^{\mu} \phi) + \nonumber \\
&+ &\frac12{\rm tr} [A_{\mu}, \phi][A^{\mu}, \phi] + V(\phi),
\end{eqnarray}
where $A_\mu$ is the total gauge field of the SM:
\be 
\label{AWBG}
A_{\mu} = i (W_{\mu}^+ I_+ + W_{\mu}^- I_- + W_{\mu}^3 I_3 + NB_\mu Y + \frac12 \sum_{s=0}^8\sum_{i,j=1,2,4}
G_\mu^s a_i^*\lambda_s^{ij}a_j),
\ee 
$W_\mu$ and $B_\mu$ are an $SU(2)_L$ triplet and singlet, respectively, $G_\mu$ is the gluon ($SU(3)_c$) octet,
$\lambda_s$  are the su(3) Gell-Mann matrices such that
$tr(\lambda_s\lambda_t) = 2 \delta_{st}$. The normalization constant $N$ is determined from the condition that $I_3$ and $NY$ are equally normalized in $J(\mathcal{P})$:
\be 
\label{NSM}
tr(I_3^L)^2(=\frac12 (1+3))= 2 = tr(NY)^2 = \frac{40}{3}N^2 \Rightarrow N^2=\frac{3}{20}.
\ee
Here we have used the calculation: $trY^2=1\times 2 + \frac19\times 6 + 4 + \frac{4}{9} \times 3 + \frac{16}{9}\times 3 = \frac{40}{3}.$
Clearly, the value of N depends on the spectrum of fundamental fermions. For the leptonic (electroweak) model one has a smaller ratio, $N^2 = \frac{1}{12}$. We shall see that the resulting $N^2$ gives the value of the computed Weinberg angle.

We will obtain the (quadratic) mass form for the electroweak gauge fields, 
\be 
\label{QWB}
\mathcal{Q}(W, B) := -\frac12 tr[W^+ I_+^L +W^-I_-^L + W^3
I_3^L + NBY, \phi]^2,
\ee
by noting that $[G, \phi] = 0$ and substituting in the third term of the Lagrangian (\ref{LAphi}) the components of $\phi(x)$ by  constant values which minimize $V(\phi)$: 
\be 
\label{phiMin}
|\phi_\alpha|^2 = \rho_\alpha, \, \alpha=0, 8, \, \rho_0+\rho_8 = m^2, \, \rho_0 \rho_8 = 0.
\ee
In writing down (\ref{QWB}) (and later) we are omitting the (contracted) vector index $\mu$ of the gauge fields. Taking further into account the relations 
\be 
\label{WpmPhi}
[W^+a_8^*a_0+W^-a_0^*a_8, \phi]^2 = [W^+(\phi_0a_8^* - \bar{\phi}_8 a_0), W^-(\phi_8a_0^*-\bar{\phi}_0 a_8)]_+,
\ee
\begin{eqnarray}
\label{W3Bphi}
[W_3I_3+NBY, \phi]^2 = \frac14(W_3+2NB)^2(\phi_0a_0^*-\bar{\phi}_0a_0)^2 \nonumber \\
+ \frac14(W_3-2NB)^2(\phi_8 a_8^* - \bar{\phi}_8 a_8)^2,
\end{eqnarray} 
and inserting the values (\ref{phiMin}) of $\phi_0, \phi_8$ that minimize the potential, we find
\begin{eqnarray}
\label{eq411}
\mathcal{Q}(W,B)=\frac14 tr\{(\rho_0 + \rho_8)(W^+ W^- + W^- W^+) \, \, \, \, \nonumber \\ + \frac12 \left(\rho_0(W_3 + 2NB)^2 + \rho_8 (W_3 -2NB)^2 \right)\} \nonumber \\ 
= 4m^2 \left(W^+W^- + W^-W^+ + \frac12(W_3^2 + 4N^2B^2)+2NBW_3 \, \varepsilon\right), \,
\nonumber \\ \varepsilon= \varepsilon(\rho_0, \rho_8) = \frac{\rho_0-\rho_8}{\rho_0+\rho_8} = \pm 1. 
\end{eqnarray}
Eq. (\ref{eq411}) tells us that the parameter $2m$ appears as the mass of the charged, $W^{\pm}$, bosons. The mixing matrix for the neutral gauge bosons $W_3$ and $B$,
$$
\begin{pmatrix}
1 & 2N\varepsilon  \\
2N\varepsilon  & 4N^2 \end{pmatrix},
$$
has determinant $0$ for $\varepsilon^2 = 1$ as ensured by the last equation (\ref{phiMin}). This implies the existence of a zero mass photon. The physical neutral gauge fields $A^\gamma$ and the $Z$-boson diagonalize the mixing matrix by a rotation on the Weinberg angle:
$$
A^{\gamma} = cB - \varepsilon s W_3, , \ Z = \varepsilon s B + c W_3,
$$
\be
\label{eq325}
c^2 = \cos^2 \theta_w = \frac1{1+4N^2}=\frac{5}{8} \, , \ s^2 = \sin^2 \theta_w = \frac{4N^2}{1+4N^2}=\frac{3}{8} \, ,
\ee
for $4N^2 = \frac35$, (\ref{NSM}). The relations (\ref{eq325}) just reflect the fermion spectrum:
\be 
\label{2ntanWei}
tg^2\theta_w = 4N^2 = \frac{tr_{J(\mathcal{P})}(2I_3^L)^2}{tr_{J(\mathcal{P})} Y^2}(=\frac35).
\ee
No wonder that the same result is derived in grand unified theories. For $4N^2=\frac13$, the value in the leptonic model based on $J_2^4$, we would have reproduced the result  $s^2=\frac14$ of \cite{roe:1999} (also obtained in  \cite{coq:1991} and earlier, under different premises, in work of Neeman and Fairley, cited in \cite{roe:1999}).

The constant $\kappa$ in $V(\phi)$ (\ref{V(Phi)}) does not appear in the mass matrix for the gauge bosons. It does affect, however, the mass square of the Higgs mass identified as the coefficient $8m^2(1+\kappa^2)$ to $\phi\bar{\phi}$ in the quadratic term of $V(\phi)$ giving
\be 
\label{mH:mW}
m_H^2 = 2(1+\kappa^2)m_w^2.
\ee
This allows to accomodate the observed relation $16m_h \approx 25m_w$ for $\kappa \lessapprox \frac12$.
 
\smallskip
We end with two remarks placing our result in a more familiar context.

\medskip
1. The Lagrangian (\ref{LAphi}) involves no coupling constants. A way to introduce the gauge coupling $g$ of the charged $W$-bosons and the gluons consists in replacing $L(A,\phi)$ (\ref{LAphi}) by $\frac1{g^2} \, L(g A , g\phi)$, a scaling that preserves the kinetic (and, more generally, the quadratic) term (cf. \cite{roe:1999}); we then identify (a multiple of) $g$ with the $W$ and $G$ gauge coupling. The couplings $g'$ of the $Z$ boson and $e$ of the photon $A^{\gamma}$ are determined by $g$ and the Weinberg angle:
\be
\label{couplings}
g' = g \, {\rm tg} \, \theta_w \, , \qquad e^2 = g^2 \sin^2 \theta_w \, , 
\ee
yielding in our case $g^2 = \frac53 \, g'^2 = \frac83 \, e^2$.

\medskip
2. Our calculation (as well as that of  \cite{roe:1999} and in the work cited there) is classical, corresponding to a tree quantum field theoretic approximation. According to the renormalization group analysis the coupling constants $g,g',\ldots$ depend on the energy scale (or the momentum transfer -- a dependance now confirmed experimentally).  Our argument, or a similar one in a grand unified theory, is believed to be exact at ``unification scale'' (at inaccessibly high energy -- up to $10^{15} - 10^{16} \, GeV$). The measured value of $\sin^2 \theta_w$ is $0.2312$ (at momentum transfer $91.4 \, \frac{GeV}{c}$). The value $\sin^2 \theta_w = \frac14$ based on the $U(2)$ electroweak theory is, in fact, closer to it than the value $3/8$ computed for the full SM.
%%%
\section{Outlook}\label{sec5}
\setcounter{equation}{0}
The fact that the euclidean extensions of the spin factors $J_2^4$ and $J_2^8$ are related to the ``structure Clifford algebras'' ${C\ell} (5,1)$ and ${C\ell} (9,1)$ makes the superconnection approach of \cite{qui:1985}, \cite{mat-qui:1986}, adopted by physicists and neatly formulated in \cite{roe:1999}, particularly natural. The generators $\Gamma_a$ of ${C\ell} (4n+1,1)$ ($n=1,2$) anticommute with the chirality operator $\gamma = \omega_{4n+1,1}$ and intertwine between the (internal symmetry counterpart of) left and right chiral fermions. This begs to identify the (multicomponent) scalar field
\be
\label{eq41}
\widehat\Phi (x) = \sum_a \phi^a (x) \Gamma_a \, , \quad \mbox{or rather} \quad {\mathcal P} \widehat\Phi (x) {\mathcal P}
\ee
where ${\mathcal P}$ projects on the particle subspace (excluding antiparticles) with the matrix valued odd part of the superconnection associated with the Higgs field.
\smallskip
The detailed explicit calculation of Sects.~3, 4 aimed to demonstrate the accessibility and the relative simplicity of this approach. \\

The inclusion of a fermionic term into the Lagrangian involves some subtleties and will be dealt with in future work. 

Let us make some comments on the description of the theory of fundamental particles of matter for one generation of the Standard Model given here. One has an internal quantum space which corresponds to the Jordan algebra $J^8_2=JSpin_9$  of hermitian $2\times 2$-matrices over $\mathbb O$ acted by the subgroup of automorphisms preserving the splitting $\mathbb O=\mathbb C \oplus \mathbb C^3$ which is the subgroup $G_{SM}$ (\ref{eq12}) of $Aut(J^8_2)=Spin(9,0)$. One also has an external classical space which corresponds to the algebra $\calc(M)$ of real functions on spacetime acted by the subgroup of automorphisms preserving the Minkowskian structure which is the Poincar\'e group. Particles are then described by modules over $J^8_2$ and $\calc(M)$ respectively that is by the Clifford algebra $C\ell_9$ or its hermitian part for the internal structure and by the module $\cals$ of sections of the (Weyl) spin bundles for the external structure. These modules being equivariant respectively by $G_{SM}$ and by the Poincar\'e group. Here, we have taken into account $C\ell_9\times \cals$ as a module over $\calc(M)$  and investigated the corresponding (super-)gauge theory. This is not the only possibility. Indeed,  from the very beginning $C\ell_9\times \cals$ is a module over the Jordan algebra $J^8_2\times \calc(M)=\calc(M,J^8_2)$. In \cite{mdv:2016b} differential calculi over general Jordan algebras and a corresponding theory of connections over Jordan modules have been defined,  which has been further developed in \cite{car-dab-mdv:2019}. Thus it would be more natural to write an action for the theory of (super-)connections over the Jordan algebra $\calc(M,J^8_2)$ (cf. the approach of \cite{mdv-ker-mad:1988b},  \cite{mdv:1991} and \cite{mdv:2001}). If one does that, a lot of supplementary scalar fields appear, namely the components of the connection in the quantum directions (i.e. over the part $J^8_2$). It is an open problem to classify these fields and to analyse their relevance for physics.

\section*{Appendix: \bf The splitting $\mathbb O= \mathbb C \oplus \mathbb C^3
$ and the associated $\mathbb Z_3$-symmetry}

 The splitting $\mathbb O=\mathbb C\oplus \mathbb C^3$ corresponds to the choice of an imaginary unit ${\mathbf i} \in\mathbb O$ which plays the role of the complex imaginary $i\in \mathbb C$. One can then write an octonion $x\in \mathbb O$ as
\[
x=z+\sum_k Z^k {\mathbf e}_k=z+Z
 \]
 where $z$ and the $Z^k$ are elements of $\mathbb C=\mathbb R+{\mathbf i}\mathbb R (\subset \mathbb O)$ and where $({\mathbf e}_k)$ is the canonical basis of $\mathbb C^3$ , $k\in \{1,2,3\}$. One recovers the product of $\mathbb O$ by setting
\[
 \left\{
 \begin{array}{l}
{\mathbf i}^2= -1\\
{\mathbf i} {\mathbf e}_k=-{\mathbf e}_k{\mathbf i}\\
 {\mathbf e}_k {\mathbf e}_\ell =- \delta_{k\ell} 1 + \sum_m \varepsilon_{k\ell m}{\mathbf e}_m
 \end{array}
  \right.
\]
i.e. the $
{\mathbf e}_k$ generate a quaternionic subalgebra. The subgroup of $G_2=\Aut(\mathbb O)$ which preserves ${\mathbf i} \in \mathbb O$ is isomorphic to $SU(3)$ and is identified in our picture to the colour group $SU(3)_c (\subset G_2)$ while the splitting $\mathbb O=\mathbb C\oplus \mathbb C^3$ is identified to the quark-lepton symmetry, $\mathbb C^3$ for the quark and $\mathbb C$ for the lepton.\\
 
 Following \cite{yok:2009}, let us consider the center $\mathbb Z_3$ of $SU(3)_c$, this is the subgroup of $G_2$ induced by the action $w$ of 
 ${\mathbf j}=-\frac{1}{2}+\frac{\sqrt{3}}{2}{\mathbf i}\in \mathbb O$ on $x=z+Z\in \mathbb O$ as
\[
 w(x)=w(z+Z)=z+{\mathbf j}Z
\]
 where $Z=(Z^k)\in \mathbb C^3\subset \mathbb O$ and ${\mathbf j}Z=({\mathbf j}Z^k)$ is the diagonal action. Then, by construction $w\in G_2$ and the subgroup of $G_2$ which commutes with $w$ is again $SU(3)_c\subset G_2$.\\
 
 Consider the Jordan algebra $J^8_2=JSpin_9=\calh_2(\mathbb O)$ of the hermitian octonionic $2\times 2$ matrices. The group of automorphisms of $J^8_2$ is the group $Spin(9)$ and the mapping
 \[
 w_2:\left(\begin{array}{cc}
 \lambda_1 & x\\
 \bar x & \lambda_2
 \end{array}
 \right)
\mapsto
\left(
\begin{array}{cc}
\lambda_1 & w(x)\\	
\overline{w(x)} & \lambda_2
\end{array}
\right)
\]
defines an automorphism of $J^8_2$ which induces an action of $\mathbb Z_3$ on $J^8_2$. The subgroup of $\Aut(J^8_2)=Spin(9)$ which commutes with this action (i.e. with $w_2$) is the group $G_{SM}$ defined by (\ref{eq12}) which preserves the splitting $\mathbb O=\mathbb C\oplus \mathbb C^3$ and the $\mathbb C$-linearity in $\mathbb C^3$.\\

Consider now the exceptional Jordan algebra $J^8_3=\calh_3(\mathbb O)$, then the mapping
\[
w_3 :
\left(
\begin{array}{ccc}
\lambda_1 & x_3 & \bar x_2\\
\bar x_3 & \lambda_2 & x_1\\
x_2 & \bar x_1 & \lambda_3
\end{array}
\right)
\mapsto
\left(
\begin{array}{ccc}
\lambda_1 & w(x_3) & \overline{w(x_2)}\\
\overline{w(x_3)} & \lambda_2 & w(x_1)\\
w(x_2) & \overline{w(x_1)} & \lambda_3
\end{array}\right)
\]
defines an automorphism of $J^8_3$ (i.e. $w_3\in F_4=\Aut(J^8_3)$) and induces an action of $\mathbb Z_3$ on $J^8_3$. The subgroup of $F_4$ which commutes with this action (i.e. with $w_3$) is the subgroup of $\Aut(J^8_3)=F_4$ isomorphic to
\[
SU(3) \times SU(3)/\mathbb Z_3 
\]
which preserves the splitting $\mathbb O=\mathbb C\oplus \mathbb C^3$ and the $\mathbb C$-linearity in $\mathbb C^3$, \cite{yok:2009}. This subgroup was denoted as $SU(3)_c\times SU(3)_{ew}/\mathbb Z_3$ in \cite{mdv-tod:2019}.\\

\noindent {\bf Warning} : 
Our presentation of $\mathbb O$ at the beginning of this appendix is clearly related to the Cayley-Dickson construction applied to the transition from $\mathbb H$ to $\mathbb O$ by adding the ``new" imaginary unit ${\mathbf i}$,  but this ${\mathbf i}\in \mathbb O$ should not be confused with the complex number $i$ involved in the complexification $C\ell_9$ of $C\ell(9,0)$  in \cite{mdv-tod:2019} and in $C\ell(10,\mathbb C)$ in Section 3.  \\

Once one works in $\mathbb O$, it is much more natural to index a basis of the imaginary octonionic units by the field $\mathbb Z_7$ of the integers modulo 7. Among such a choice the choice of \cite{bae:2002} is particularly nice since in the basis $(e_\alpha)_{\alpha \in\mathbb Z_7}$ of 
\cite{bae:2002} the relations of $\mathbb O$ (i.e. the multiplication table of $\mathbb O$) are translational invariant
\[
e_\alpha e_\beta = e_\gamma \Rightarrow e_{\alpha+1} e_{\beta+1} = e_{\gamma+1}
\]
and invariant by the dilatation by 2, i.e. 
\[
e_\alpha e _\beta = e_\gamma\Rightarrow e_{2\alpha} e_{2\beta} = e_{2\gamma}
\] 
so that everything is fixed by setting $e_1e_2=e_4$ (which is then necessary for the consistence) and we stick to the above choice  for $\mathbb O$. In such a basis $e_7(``=e_0")$ has the particularity to be invariant by dilatation
\[
e_{\alpha 7} = e_7,\> \forall \alpha\in \mathbb Z_7
\]
and is unique under this condition since 7 is a prime number (i.e.$\mathbb Z_7$ is a field).\\

Since in our approach the splitting $\mathbb O=\mathbb C\oplus \mathbb C^3$ is fundamentally linked to the color symmetry of quarks and to the quark-lepton symmetry \cite{mdv:2016b}, it is natural to identify ${\mathbf i}\in \mathbb O$ as ${\mathbf i}=e_7$ $(``=e_0")$ in this frame.
This justifies our choice of notations all along our paper. The relation ${\mathbf i}=e_7$ must be supplemented by $\
{\mathbf e}_1=e_1,  {\mathbf e}_2=e_2$ and $
{\mathbf e}_3=e_4$ to express the previous items in term of basis $(e_\alpha)_{\alpha\in\mathbb Z_7}$ of \cite{bae:2002}.

\bigskip

\bigskip

\noindent {\bf Acknowledgments}

\medskip
The authors thank the referee for his careful reading and constructive remarks taken into account in the final version of the paper. I.T. thanks IHES for hospitality during the course of this work. It is a pleasure to thank C\'ecile Gourgues for her kind help and expert typing. The help of Svetla Drenska and Ludmil Hadjiivanov in revising earlier drafts is gratefully acknowledged. I.T. acknowledges the hospitality for a brief period at the Theory Department of CERN and thanks Anton Alekseev and the Section de Math\'ematiques, Universit\'e de Gen\`eve for invitation and support during the final stage of the work.   

\newpage

%\bibliographystyle{plain}
%\bibliography{BibMich,BibExt}

\end{document}